\documentstyle[12pt, epsfig]{article}

\begin{document}

\author{A. Guarino$^{{\sf *}}$, S. Ciliberto$^{{\sf *}}$ and A. Garcimart\'{i}n$^{%
{\sf \ddagger }}$ \\
$^{{\sf *}}${\small Ecole Normale Superieure de Lyon, 46 all\'{e}e d'Italie,
69364 Lyon, France}\\
$^{{\sf \ddagger }}${\small Departamento de F\'{i}sica, Facultad de
Ciencias, Universidad de Navarra,}\\
{\small \ E-31080 Pamplona, Spain.}}
\title{Failure time and microcrack nucleation }
\date{}
\maketitle

\begin{abstract}
The failure time of samples of heterogeneous materials (wood, fiberglass) is
studied as a function of the applied stress. It is shown that in these
materials the failure time is predicted with a good accuracy by a model of
microcrack nucleation proposed by Pomeau. It is also shown that the crack
growth process presents critical features when the failure time is
approached.
\end{abstract}

\bigskip

\bigskip


\vspace{2cm}

It is very well known that different materials subjected to a constant load
may break after a certain time, which is a function of the applied load
[1-5]. Many models have been proposed to predict this failure time, but the
physical mechanisms remain often unclear \cite{sha, libro}. Very recently
Pomeau proposed a model \cite{pom}, which explains quite well the failure
time of microcrystals \cite{pauch} and gels \cite{bonn} submitted to a
constant stress. This model is based on the interesting idea that a
nucleation process of microcracks has to take place inside the materials, in
order to form the macroscopic crack. This nucleation process is controlled
by an activation law, as the coalescence of a phase into another in a
liquid-solid transition. Based on this prediction \cite{pom}, L. Pauchard et
al. \cite{pauch} found that if a constant load is applied to a bidimensional
microcrystal, it breaks after a time $\tau $ given by the equation $\tau
=\tau _oe^{P_o^2/P^2}$, where $P$ is the applied pressure, and $\tau _o$ and
$P_o$ are constants. Bonn et al. \cite{bonn} found a similar law for gels.
Pomeau predicted that for three-dimensional microscopic systems the
life-time should be

\begin{equation}
\tau =\tau _oe^{\left( P_o/P\right) ^4}  \label{const}
\end{equation}

where $\tau _o$ is a characteristic time and $P_o$ a characteristic
pressure, which mainly depend on the material characteristics, the
experimental geometry and temperature. This idea is quite interesting and it
merits to be checked experimentally in hetrogeneous materials, such as fiber
glass and wood pannels. Indeed, in two recent papers \cite{prl, articolo},
we have shown that in these materials the microcracks, preceding the main
crack form something like a coalescence around the final path of the main
crack.

Driven by these observation we decided to study the behaviour of these
materials as a function of time and to check whether eq.1 could be useful in
order to predict the sample life time. To do this we monitor the acoustic
emission (AE) released before the final break-up by a sample placed between
two chambers between which a pressure difference $P$ is imposed. In Fig. 1 a
sketch of the apparatus is shown. We have prepared circular wood and
fiberglass samples with a diameter of 22 cm. and a thickness of 4 mm (wood)
and 2 mm (fiberglass). In our samples the AE consists of ultrasound bursts
(events) produced by the formation of microcracks inside the sample. For
each AE event, we record the energy $\varepsilon $ detected by the four
microphones, the place where it was originated, the time at which the event
was detected and the instantaneous pressure and deformation of the sample.
The energy is defined as the integral of the sum of the squared signals. A
more detailed description of the experimental methods can be found in \cite
{prl,articolo}.

We first investigate the behaviour of the samples as a function of time when
they are submitted to a constant load. Our interest is focused on the
life-time of the sample, on the behaviour of the released acoustic energy
near the fracture and on the distributions of energy and time elapsed
between two consecutive events. The behaviour of the energy as a function of
time for a system submitted to a constant load has been studied by
geologists, but they were not specially interested in what happened near the
fracture \cite{mogi,stead,maes}.

We first imposed a constant strain to our samples, as it has been made for
crystals. As strain is fixed, every microcrack leads to a pressure decrease,
so the system reaches a stationary state. This is because a microfracture
weakens the material. In the absence of microcraks the pressure remains
constant. One sample was submitted to a large deformation (close to the
fracture) and it did not break after three days. Therefore, at imposed
strain, the effect observed in microcrystals is not valid for heterogeneous
materials. On the other hand, if a constant stress is applied to the system,
it will break after a certain time which depends on the value of the applied
pressure. The reason for this is that after every single microcrack the same
load must be endured by the weakened sample, so it becomes more and more
unstable. We have submitted several samples to different constant pressures
and we have measured the time until the break-up (the life-time $\tau $).
The values obtained are well fitted by eq.(\ref{const}), that is the
exponential function predicted by Pomeau; the life-time expression $\tau
=ae^{-bP}$ proposed by Mogi \cite{mogi}, on the other hand, does not conform
to our data\cite{commento}. In Fig. 2a $\tau $ is plotted versus $\frac
1{P^4}$ in a semilog scale, and a straight line is obtained. Even if the
pressure difference is very small the sample will eventually break, although
the life-time can be extremely long. For example, using eq. (\ref{const})
and the best fit parameters of fig. 2a, one estimate  $\tau \simeq 5000$ s
at $P=0.43$ atm. Halving the imposed pressure causes $\tau $ to become
extremely large : $\tau =4.4\cdot 10^{37}$ years at $P=0.21atm$).

When a constant pressure is applied to the sample, the acoustic emission of
the material is measured as a function of time. We find that the cumulated
acoustic energy $E$ diverges as a function of the reduced time $\frac{\tau-t%
}{\tau}$, specifically $E \propto (\frac{\tau- t}{\tau})^\gamma$ with $%
\gamma=0.27$ (see Fig. 2b). Notably, the exponent $\gamma$, found in this
experiment with a constant applied pressure, is the same than the one
corresponding to the case of constant stress rate \cite{prl}. Indeed it has
been shown \cite{prl,articolo} that if a quasi-static constant pressure rate
is imposed, that is $P=A_pt$, the sample breaks at a critical pressure $P_c$
and $E$ realesed by the final crack precursors (microcrcaks) scales with the
reduced pressure or time (time and pressure are proportional) in the
following way:

\begin{equation}  \label{Energy}
E \propto (\frac{P_c-P}{P_c})^\gamma= (\frac{\tau-t}{\tau})^\gamma
\end{equation}

where $\tau=P_c/A_p$ in this case. Thus it seems that the real control
parameter of the failure process is time, regardless of the fact that either
a constant pressure rate or a constant pressure is applied.

To find a general law, which is valid for a time dependent imposed stress,
we intend to generalize the eq.\ref{const} which is valid only for a
constant imposed pressure. In the case where the pressure changes with time,
it is reasonable to consider the entire history of the load. Therefore we
consider that
\[
{\frac 1{\tau _o}}\exp (-(\frac{P_o}P)^4)
\]
is the density of demage per unit time. The certitude of breaking is
obtained after a time $\tau $ such that:
\begin{equation}
\int_0^\tau \frac 1{\tau _o}e^{-(\frac{P_o}P)^4}dt=1  \label{integ}
\end{equation}
where $\tau _o$ and $P_o$ have the previously determined value. Notice that
this equation is equivalent to eq. \ref{const} when a constant pressure is
applied.

To test this, we have applied the load to the sample following different
schemes. We have first applied successive pressure plateaux in order to
check whether memory effects exist. In fig.3a the pressure applied to the
sample is shown as a function of time. A constant load has been applied
during a certain time $\tau _1$, then the load is suppressed and then the
same constant load for a time interval $\tau _2$ is applied again.The sample
breaks after a loading a time $\tau _1+\tau _2$ which is equal to the time
needed if the same load had been applied continuosly without the absence of
load during a certain interval. Therefore there is a memory of the load
history. The life-time formula (eq. \ref{integ}) is also valid if different
constant loads are applied successively (fig. 3b). This concept can explain
the violation of the Kaiser effect in these materials \cite{articolo}.

If the load is not constant, the life-times resulting from the proposed
integral equation are still in good agreement with experimental data. A load
linearly increasing at different rates $A_p$ has been applied to different
samples. The measured breaking times are plotted in fig.4 along with a curve
showing the values computed from eq.\ref{integ}. Even if a quasi-static load
is applied erratically (fig. 3d), the calculated life-time agrees with the
measured one. These experiments show that eq. \ref{integ} describes well the
life-time of the samples submitted to a time dependent pressure.

The question is to understand why eq.(\ref{const}) and (\ref{integ}) works
so well for a three dimensional heterogeneous material. Indeed, in the
Pomeau formulation
\begin{equation}
P_o=G \left( \frac{\eta ^3Y^2}{kT}\right) ^{1/4}  \label{PO}
\end{equation}
where Y is the Young modulus,T the temperature, K the Boltzmann constant and
$\eta $ the surface energy of the material under study. G is a geometrical
factor which may depend on the experimental geometry, on defect shape and
density.

In our experiment, we found $P_o=0.62$ atm for wood, which has Y=$1.8\cdot
10^8$ N/m$^2$, and $P_o=2.91$ atm for fiberglass, which has Y=$10^{10}$ N/m$%
^2$. Thus the ratio between the values of $P_o$ found for the two materials
is closed to the ratio of the square root of their Young modula.

In contrast temperature does not seem to have a strong influence on $\tau $.
In fact we changed temperature, from $300K$ to $380K$ which is a temperature
range where the other parameters, $Y$ and $\eta $, do not change too much.
For this temperature jump one would expect a change in $\tau $ of of about $%
50\%$ for the smallest pressure and of about $100\%$ for the largest
pressure. Looking at fig.4 we do not notice any change of $\tau $ within
experimental errors which are about $10\%$. In order to maintain the change
of $\tau $ within $10\%$ for a temperature jump of $80K$ one has to assume
that the effective temperature of the system is about $3000K$. Notice that
this claim is independent on the exact value of the other parameters and G.

These observations seem to indicate that the nucleation process of
microcraks is activated by a noise much larger than the thermal one. Such a
large noise can be probably produced by the internal random distribution of
the defects in the heterogeneous materials that we used in our experiments.
This internal random distribution of material defects evolves in time
because of the appearance of new microcracks and the deformation of the
sample. Therefore this internal and time dependent disorder of the material
could actually be the mechanism that activates the microcrak coalescence and
play the role of a very high temperature. Similar conclusions about a
disordered induced high temperature have been reached in other disordered
systems \cite{sollich}. This is an important point that merits to be deeply
explored. Simple numerical simulations which we performed in fuse networks
seem to confirm this hypothesis.

As a conclusion we have shown that a model based on the nucleation of
microcracks is in agreement with experimental data of failure of two
heterogeneous materials. This model seems to be quite general because it
also explains the failure of gels \cite{bonn} and microcrystals \cite{pauch}%
. It will be certainly interesting to test it in other materials. However
many questions remain open. The first one concerns the high temperature of
the system. The second is related to the behaviour of acoustic energy close
to the failure time. Indeed if AE is considered as a susceptibility \cite
{zapperi} it is not easy to put together the observed critical divergency
with a nucleation process. Probably the standard phase transition
description can be only partially applied to failure because of the
intrinsic irreversibility of the crack formation. All of these are very
interesting aspects of this problem which certainly merit to be clarified in
the future.

Correspondence should be addressed to S. Ciliberto (e-mail:
cilibe@physique.ens- lyon.fr) \newpage

{\bf Figure Captions}

\smallskip

\begin{enumerate}
\item  Sketch of the apparatus. S is the sample, DS is the inductive
displacement sensor (which has a sensitivity of the order of 1 $\mu $m). M
are the four wide-band piezoelectric microphones. P=P$_1$-P$_2$ is the
pressure supported by the sample. P is measured by a differential pressure
sensor ( sensitivity = 0.002 atm) that is not represented here. EV is the
electronic valve which controls P via the feedback control system Ctrl. HPR
is the high-pressure air reservoir.

\item  Measures on wood samples: {\bf a)} The time $\tau $ needed to break
the wood samples under an imposed constant pressure P is here plotted as a
function of 1/P$^4$ in a semilog scale. The dashed line represents the
solution proposed by Mogi \cite{mogi}($\tau =ae^{-bP}$). The continuous line
is the solution proposed by Pomeau for microcrystals ($\tau =\tau
_oe^{(P_o/P)^4}$). In the plot $\tau _o=50.5$ s and $P_o=0.63$ atm. Every
point is the average of 10 samples. The error bar is the statistical
uncertainty. For the fiberglass samples, we find $\tau _o=44.6$ s and $%
P_o=2.91$ atm. {\bf b)} The cumulated energy E, normalized to E$_{\max }$,
as a function of the reduced control parameter $\frac{\tau -t}\tau $ at the
neighborhood of the fracture point (Case of imposed constant pressure). The
circles are the average for 9 wood samples. The solid line is the fit $%
E=E_0\left( \frac{\tau -t}\tau \right) ^{-\gamma }$. The exponent found, $%
\gamma =0.26$, does not depend on the value of the imposed pressure. In the
case of a constant pressure rate the same law has been found \cite
{articolo,prl}.

\item  The imposed time dependent pressure (bold dotted line) is plotted as
a function of time in the case of wood samples. The continuous line is the
integral in time of the function $f\left( P\right) ={\frac 1{\tau _o}}%
e^{-P_o^4/P^4}$. On the basis of eq. \ref{integ} the predicted breaking time
$\tau $ is obtained when the integral of $f\left( P\right) $ is equal to 1.
The horizontal distance between the two vertical dashed lines in each plot
represent the difference between the predicted and the measured breaking
time. In {\bf a)} a constant pressure has been applied during about 700 s,
then the load is suppressed and then the same constant load is applied
again. The difference between the life-time predicted by (eq. \ref{integ})
and the experimental result is of $3\%$. {\bf b) } Here two pressure
plateaux of different value are successively applied to the sample. The
difference between the measured and the predicted life-time is of $5\%$. In
{\bf c)} an erratic pressure is applied to the sample. Here the error is of $%
10\%$.

\item  A load linearly increasing at different rates $A_p$ has been applied
to different samples. The measured breaking times are plotted as a function
of $A_p$ in a loglog scale; circles and squares represent the measures on
wood and fiberglass samples respectively at T=300 $K$. Bold triangles
represent measures on wood samples at T=380 $K$. The lines are the life time
calculated from eq.\ref{integ} using the best fit values for $P_o$ and $\tau
_o$. These experiments show that eq. \ref{integ} describes well the
life-time of the samples submitted to a time dependent pressure.
\end{enumerate}

\begin {figure}[p]
\psfig {figure=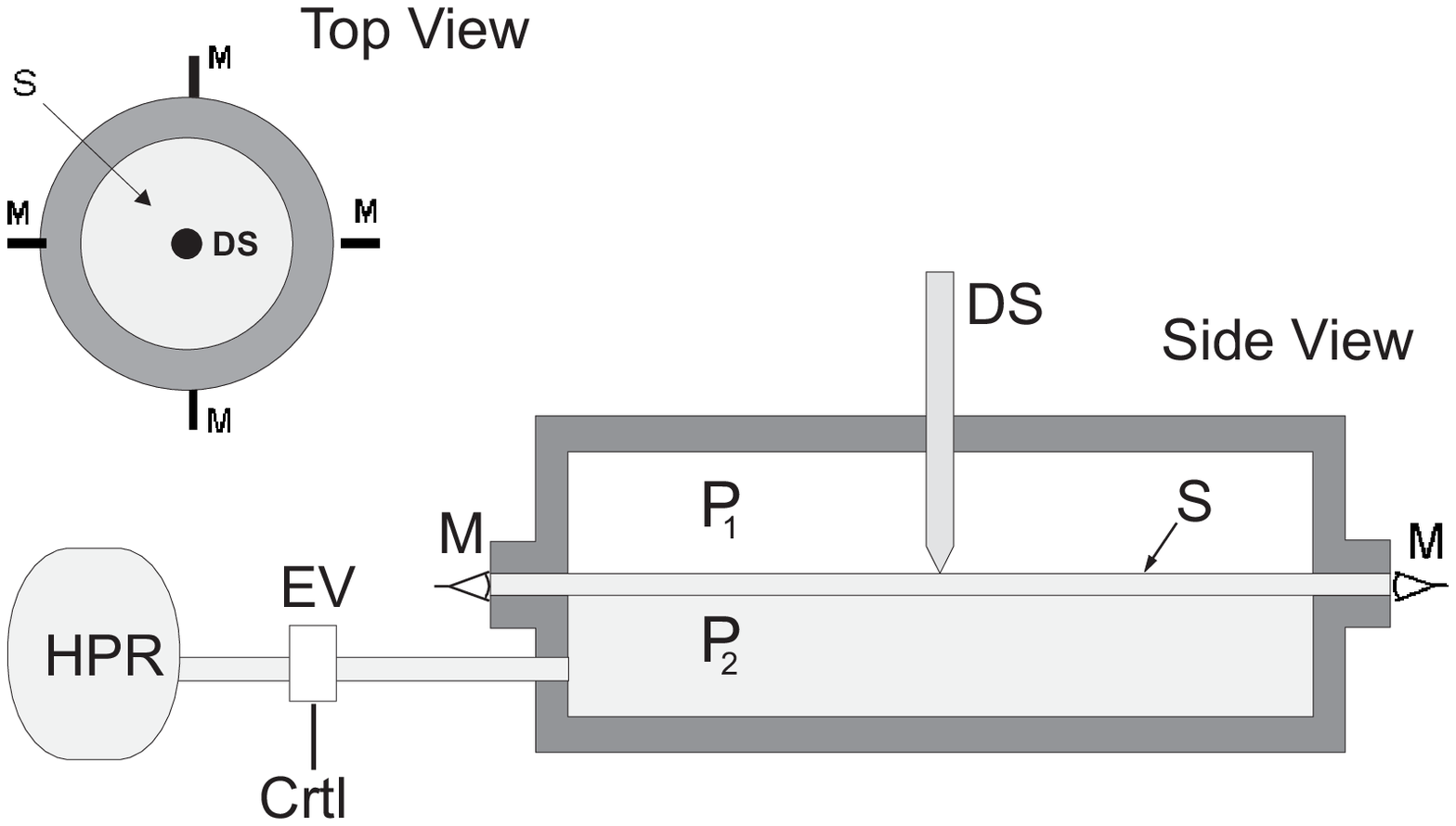}
\end {figure}

\begin {figure}[p]
\psfig {figure=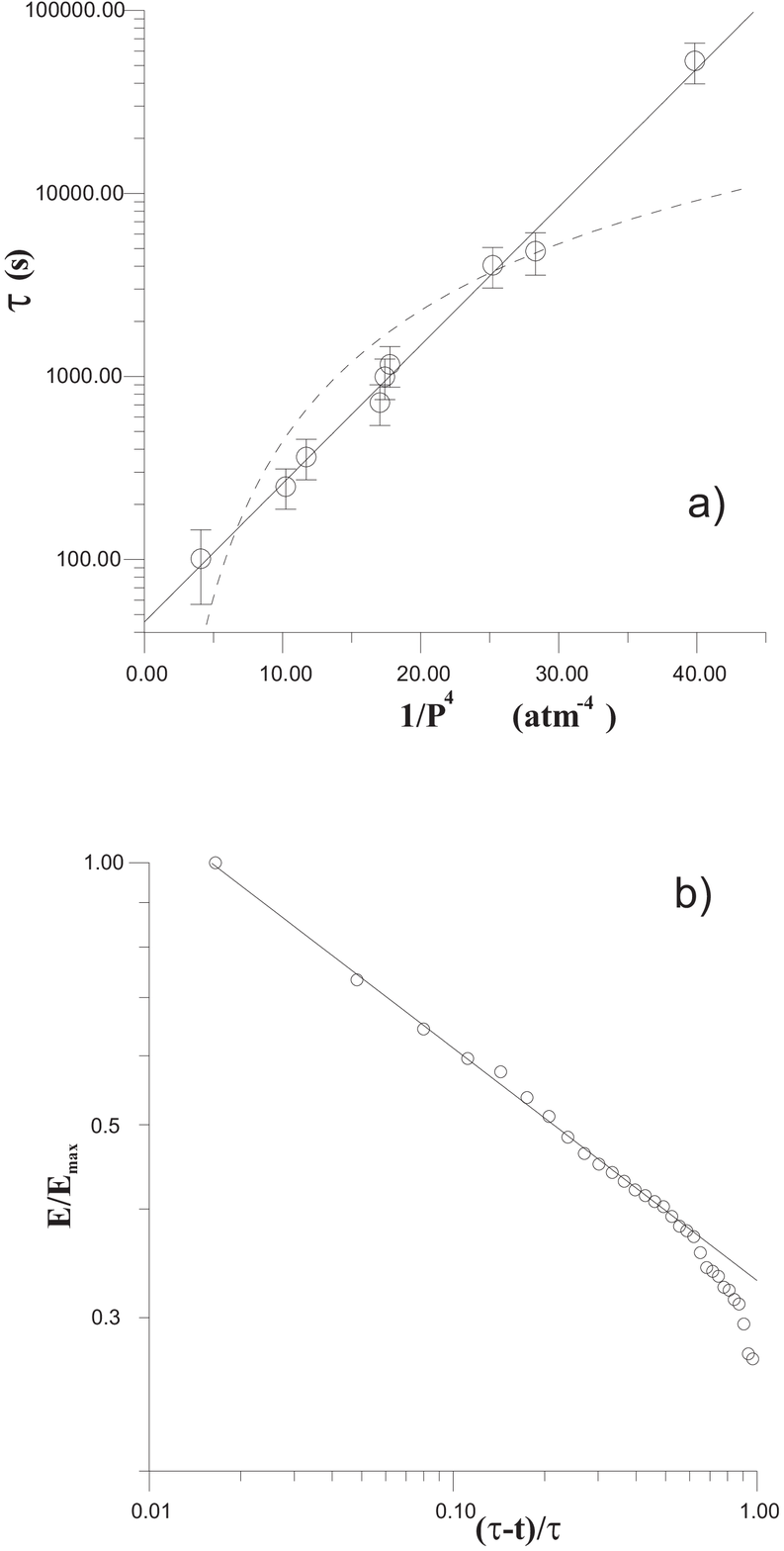, height=17cm}
\end {figure}

\begin {figure}[p]
\psfig {figure=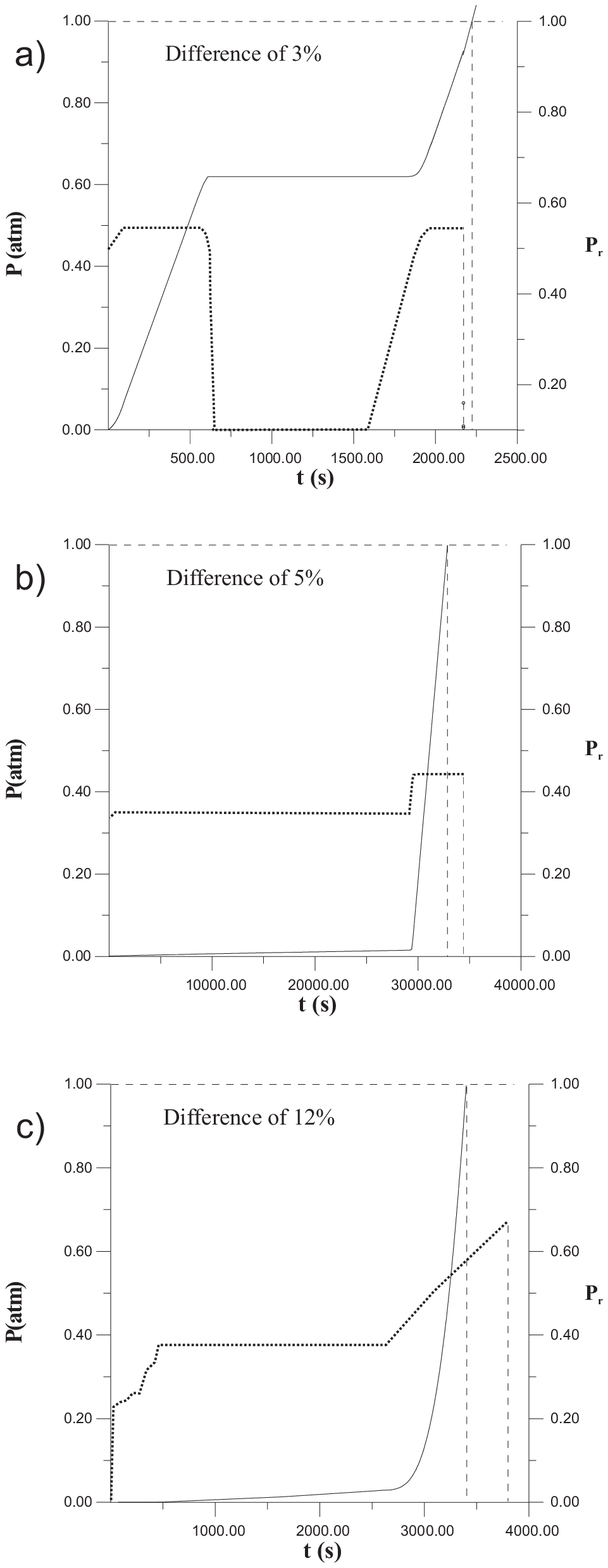, height=17cm}
\end {figure}

\begin {figure}[p]
\psfig {figure=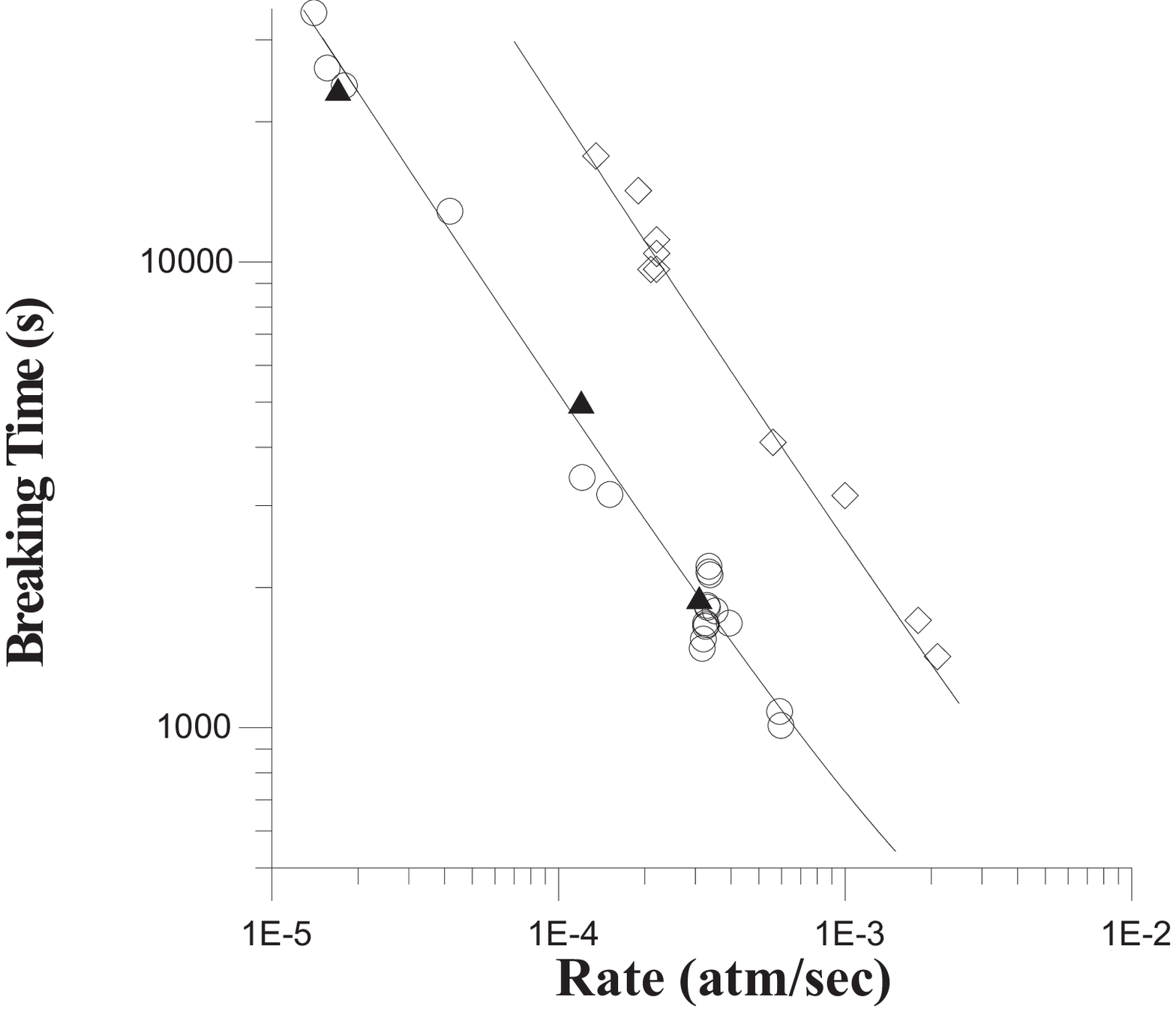, height=13cm}
\end {figure}


\begin{thebibliography}{99}
\bibitem{commento}  Several functions with many free parameters have been
proposed [1-5] in order to fit the dependence of the sample life time on the
applied stress. Among the fitting functions, which are supported by a
microscopic physical mechanism we show in fig.2a the two functions which
give the best fit of our data.
\end{thebibliography}
\end{document}